\documentclass{article}\hoffset=-2.5cm\textwidth=18cm
\usepackage{graphicx}
\newcommand{\be}{\begin{equation}}\newcommand{\ee}{\end{equation}}
\newcommand{\bea}{\begin{eqnarray}}\newcommand{\eea}{\end{eqnarray}}
\newcommand{\J}{Jarzynski }

\begin{document}

\title{On practical applicability of the \J relation in statistical mechanics: a pedagogical example}
\author{Rhonald C. Lua, Alexander Y. Grosberg \\ Department of Physics, University of Minnesota \\
116 Church Street SE, Minneapolis, MN 55455}

\maketitle

\abstract

We suggest and discuss a simple model of an ideal gas under the
piston to gain an insight into the workings of the \J identity
connecting the average exponential of the work over the
non-equilibrium trajectories with the equilibrium free energy.  We
show that the \J identity is valid for our system due to the very
rapid molecules belonging to the tail of the Maxwell distribution. For
the most interesting extreme, when the system volume is large,
while the piston is moving with large speed (compared to thermal
velocity) for a very short time, the necessary number of
independent experimental runs to obtain a reasonable approximation
for the free energy from averaging the non-equilibrium work grows
exponentially with the system size.

\section{Introduction}

The celebrated \J identity is perhaps the most recently discovered
\cite{Jar1,Jar2} simple general formula in elementary statistical
mechanics:
\be \langle e^{ W/k_B T} \rangle = e^{- \Delta F / k_B T} \
\label{eq:Jarz} \ee
(see below about sign convention).  The claim of this charming
simple formula is as follows. Suppose we have an arbitrary system
and let us consider two states of this system specified by
parameters, say $A_{\rm initial}$ and $A_{\rm final}$; these could
be volumes, magnetic fields, or just about anything else.  If the
system comes to thermodynamic equilibrium at $A_{\rm initial}$, it
has free energy $F(A_{\rm initial}) = F_{\rm initial}$; if the
system is equilibrated at $A_{\rm final}$, its free energy is
$F(A_{\rm final}) = F_{\rm final}$. The difference of these free
energies is $\Delta F = F_{\rm final} - F_{\rm initial}$.
According to elementary thermodynamics, if we drive the system
from initial to final state by a reversible process, such that the
system remains at equilibrium at every stage, then we have to
perform work, $-W$, which is equal the free energy change:
$-W = \Delta F$; if, on the other hand, the process is not
reversible, then the second law of thermodynamics tells us that $\left<-W\right>
\geq \Delta F$. We know, of course, that the second law of
thermodynamics is of a statistical nature, and, therefore, from
time to time, very infrequently, the fluctuations occur in which
$-W < \Delta F$. These fluctuations might be very rare, but with
large $W$ (strongly negative $-W$) their contribution to the
average of $e^{W/k_BT}$ might be significant. The \J formula
(\ref{eq:Jarz}) tells us precisely that: when all fluctuations,
including those violating $-W \geq \Delta F$, are taken
into account, then the average of $\langle e^{ W/k_B T} \rangle$
reduces to $e^{- \Delta F / k_B T}$.

It is worth repeating that $-W$ is the work performed by an
external force on the system; in other words, $W$ is the work
performed by the system itself.  We use this sign convention
(perhaps somewhat non-standard) because it will make for positive
$W$ and save us some writing in interesting cases below.

The tempting use of this result is to circumvent in computer
simulation or in real experiment the often painful stage of
equilibrating the system.  Instead, it should be possible to run
the system many times without any worry of the equilibrium,
repeatedly measure the work, $W$, and still obtain the equilibrium
information, $\Delta F$ from the formula  (\ref{eq:Jarz}).  The
problem is that it requires exploration of all sorts of fluctuations
along the way and measurement of the work $W$ for every fluctuation.

There are already quite a number of works exploring various aspects
of the \J formula.  Papers \cite{Crooks,Kurchan,Maes}
address mathematical foundations of the \J identity in the context of
fundamental statistical mechanics.  Some authors describe the \J identity
in terms of `transient violations of the second law of thermodynamics'
\cite{Ritort}.  There is a history of cases when such fluctuation
effects were mentioned under the name of `temporary violations of the
second law' \cite{Evans,Wang}, but many people feel that it is incorrect
terminologically to say that fluctuations violate the second law,
even if temporarily.  Here, we do not make any firm commitment to any
terminology in this sense, and just say that the \J identity is based on
proper exploration of a representative set of fluctuations.  Various
ways to apply the \J relation in computer simulations and in
experiments are discussed in
\cite{Ritort,Hummer1,Liphardt,Seifert1,Seifert2} (see also extensive
literature cited in these works).

Nevertheless, we found that one aspect is missing in the current
literature, namely, the pedagogical aspect. The \J formula appears
so simple and so general that there ought to be a simple way to
explain it and to gain an intuitive insight of it on a very
elementary level. This motivated us to look for the simplest
possible example in which the \J equation shows some non-trivial
results. Since the conceptually simplest subject in statistical
mechanics is undoubtedly the classical ideal gas, the goal of the
present paper is to work out the application of the \J formula to
the ideal gas. Of course, we shall find nothing really new in
terms of factual results, but we hope for a good new insight.

\section{Model and formulation of the seeming paradox}

Consider some amount of ideal gas in a vessel under a piston.
Everything is supposed to be in accord with elementary physics
textbooks: ideal thermoisolation, mass-less piston moving without
any friction, etc.  Suppose initially the piston is some distance
$L$ from the bottom of the vessel, and that the gas temperature is
$T$.  Let us now move the piston by some distance $\Delta L$ and
stop it again, thus preparing the final state.

The following way of thinking seems quite logical from a physicist
point of view.  To make the situation dramatic, let us suppose
that we move the piston at a very high speed, much faster than the
speed of sound in the gas, or, in other words, much faster than
the averaged thermal velocity of the molecules.  Then we can
roughly say that no molecules will be able to chase the piston
while it is moving, no molecules will hit it, and there will be,
therefore, no work.  When the piston is stopped at the end,
molecules start arriving, they do bombard the piston, but since
the piston does not move at this stage, the work is still zero.
This logic leads to the conclusion that in this case $W=0$,
implying $\langle e^{W/k_B T} \rangle = 1$, while obviously
$\Delta F \neq 0$, which seems to contradict the \J identity
(\ref{eq:Jarz}).

To resolve this paradox we have to remember about the tail of the
Maxwell distribution: however large is the speed of a piston,
there is still some probability of molecules moving fast enough to
chase the piston and hit it while it is moving.  This already
suggests that the \J identity (\ref{eq:Jarz}) has to do with the
tails of the relevant distributions.  To make this statement more
precise, we shall compute the probability distribution of the work
$W$ for our elementary model.  This is obviously much more than
just computing the average involved in \J formula (\ref{eq:Jarz}).

In order to make our article more pedagogical, we shall start with
proving the very identity (\ref{eq:Jarz}) for our specific system.
We shall also relegate cumbersome calculations to the appendix.

\section{Calculations}

\subsection{Average value of $e^{W/k_BT}$}\label{sec:average}

Since we plan to consider an ideal gas, all molecules will
contribute to both $W$ and $\Delta F$ independently.  Therefore,
we can imagine the \J formula (\ref{eq:Jarz}) re-written as
\be \left(\langle e^{ W_1/k_B T} \rangle\right)^N = \left(e^{-
\Delta F_1 / k_B T} \right)^N \ ,  \ee
where $W_1$ and $\Delta F_1$ are the work and the free energy
change per one molecule.  We see that for the ideal gas,
quantities $W_1$ and $\Delta F_1$ satisfy the \J formula looking
identical to (\ref{eq:Jarz}).  We, therefore, restrict ourselves
for simplicity to the ``ideal gas'' of just one molecule, and also
for simplicity we suppress the index $1$ in writing $W$ and $\Delta
F$. Thus, we keep considering formula (\ref{eq:Jarz}), but we
think now about just one molecule in an ideal gas.

Furthermore, to simplify writing, we assume that the temperature
is such that $k_B T = 1$, the mass of the molecule is $m=1$, and
the piston is moving during the time interval $\tau =1$.

\begin{figure}
\centerline{\scalebox{0.6}{\includegraphics{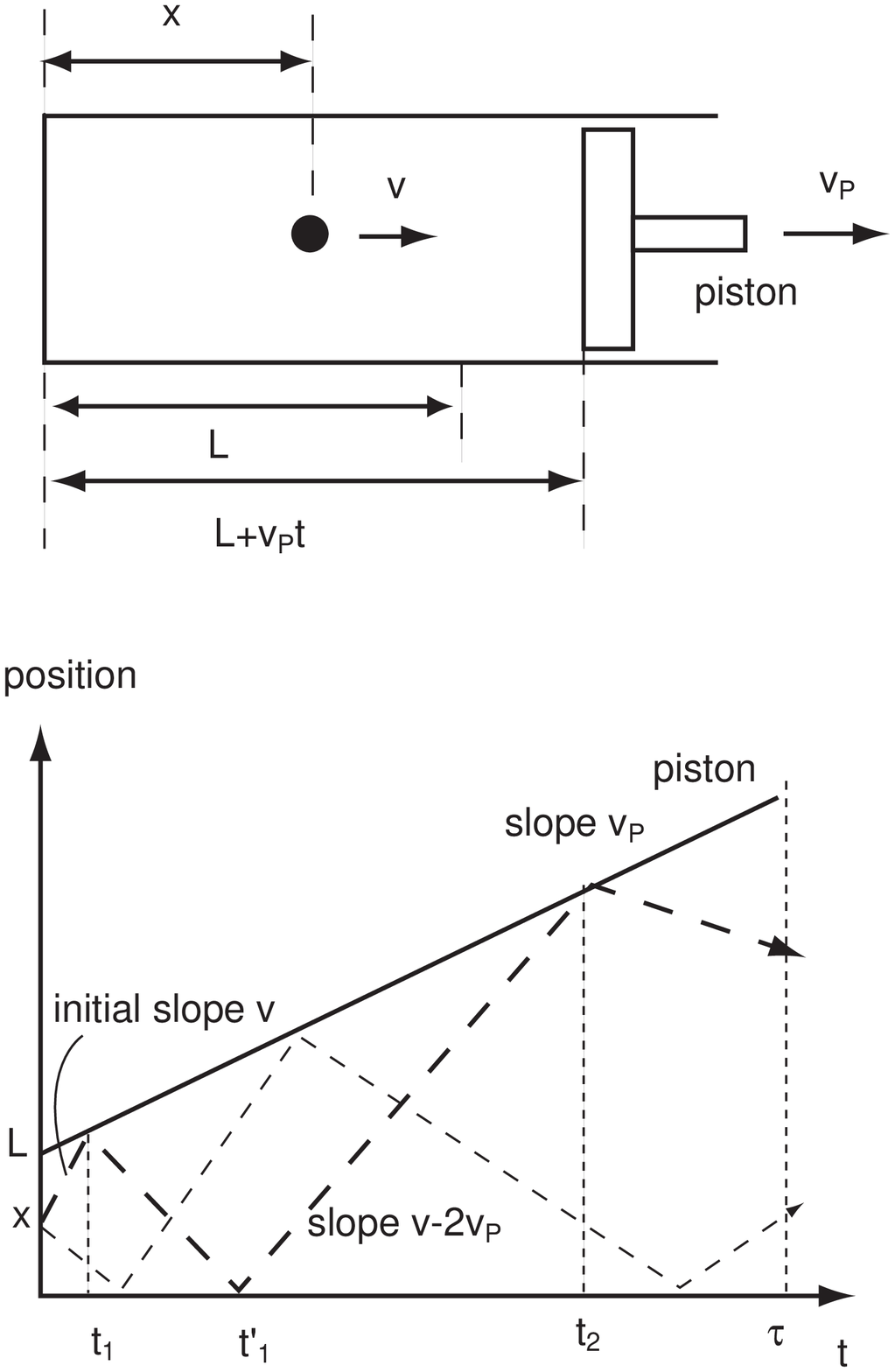}}}
\caption{System and space-time diagram. Solid line - piston
trajectory; Thick dashed line - molecule trajectory with positive
initial velocity; Thin dashed line - molecule with negative
initial velocity.} \label{fig:pistondiagram}
\end{figure}

Figure \ref{fig:pistondiagram} illustrates the system consisting
of a thermally-isolated cylinder, a piston moving at speed $v_p$,
and a single molecule initially at position $x$ with velocity $v$.
The molecule bounces off the walls elastically, so we are
concerned only with the one-dimensional motion indicated. The
space-time diagram depicts the trajectory of the piston and the
trajectories for a molecule initially moving toward the piston
(dashed lines) and for a molecule initially moving away from the
piston (thinner dashed lines).

Let us focus on the work $W$ done by the single molecule on the
piston, in a time interval $\tau =1$. The quantity $e^{W}$ ($k_B T
=1$) is to be averaged over the possible initial states $(x,v)$ of
the molecule drawn from a Maxwell-Boltzmann distribution.
Therefore,
\begin{equation}
\langle e^{W}\rangle = \frac{\int_0^Ldx\int_{-\infty}^{\infty}dv
e^{- v^2/2} e^{w_\tau(x,v)}} {\int_0^Ldx\int_{-\infty}^{\infty}dv
e^{- v^2 /2}} \label{average}
\end{equation}
where $w_\tau(x,v)$ is the work done by the gas molecule given the
initial coordinate $(x,v)$ and time elapsed $\tau$. To compute
this function $w_\tau(x,v)$, we first need to work out the
collision times between the molecule and the piston and the work
done after $n$ collisions.

Let us first assume a positive initial velocity, in which the
molecule can strike the piston first before hitting the left end
of the cylinder. The time taken for the first collision with the
piston is $t_1=\frac{L-x}{v-v_p}$. After the collision, the
velocity of the molecule relative to the piston gets reversed and
the speed of the molecule gets diminished to $v-2v_p$ (assuming
$v>2v_p$). The time taken for the second collision with the piston
is given by $t_2=\frac{3L-x}{v-3v_p}$. In general, for the
$n^{th}$ collision
\begin{equation}
t^{+}_{n}=\frac{(2n-1)L-x}{v-(2n-1)v_p}
\end{equation}
Similarly, for a molecule with a negative initial velocity,
\begin{equation}
t^{-}_{n}=\frac{(2n-1)L+x}{v-(2n-1)v_p}
\end{equation}

These relations can be inverted to give conditions that should be
satisfied by the speed of the molecule in order to result in
exactly $n$ collisions with the piston within a time interval
$\tau = 1$. For positive initial velocities,
\begin{equation}
(2n-1)(L+v_p)-x<|v|<(2n+1)(L+v_p)-x\label{eq:ineq1}
\end{equation}
For negative initial velocities,
\begin{equation}
(2n-1)(L+v_p)+x<|v|<(2n+1)(L+v_p)+x \label{eq:ineq2}
\end{equation}

The work done by the piston on the molecule after one collision is
the change in momentum of the molecule times the velocity of the
piston,
\begin{equation}
-w_1=(-(v-2v_p)-v)v_p=-2(v-v_p)v_p
\end{equation}
In general, the work done after $n$ collisions is
\begin{equation}
-w_n=-2vv_pn+2v_p^2n^2
\end{equation}
Note that the work done can also be calculated from the change in kinetic energy after $n$ collisions,
\begin{equation}
-w_n=\frac{1}{2}(v-2nv_p)^2-v^2/2=-2nv_pv+2n^2v_p^2
\label{eq:work_in_n}
\end{equation}
The work done by the molecule on the piston is positive for an
expanding volume.

We are now facing the laborious task of calculating the integral
in the numerator of formula (\ref{average}).  It is cumbersome,
because it must include the summation over all possible numbers of
bounces of our molecule from the piston.  Note that a large number
of bounces correspond to a very large initial velocity of the
molecule, as it has to have time to chase the piston for $n$ bounces,
even though it looses momentum and gets slower at every bounce.
The actual calculation is described in appendix
\ref{sec:calculating_I}.  Using the simplified result
(\ref{eq:numerator}), the sought average is
\begin{eqnarray}
\langle e^{  W} \rangle &=&\frac{\int_0^{L+v_p}dx\int_{-\infty}^{\infty}dv e^{-  v^2/2}}{\int_0^Ldx\int_{-\infty}^{\infty}dv e^{-  v^2/2}}\nonumber\\
&=&\frac{L+v_p}{L} \label{eq:Jarz1}
\end{eqnarray}
which can be recognized as the ratio of the partition functions at
the final (after time $\tau=1$) and initial volumes at the
initial temperature $T=1/ k_B$, $Z(L+v_p\tau,T)/Z(L,T)$.  This
expression is also identical to that obtained from the \J
identity.

\subsection{Probability distribution of the work $W$}

\begin{figure}
\centerline{\scalebox{0.5}{\includegraphics{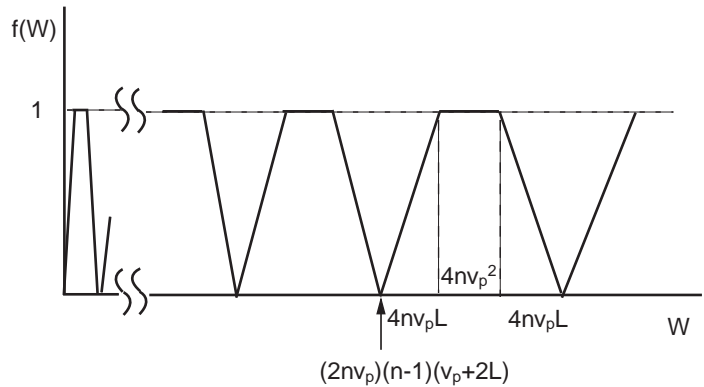}}}
\caption{The structure of the overlap factor $f(W)$, which modulates the exponential in the
distribution function. This factor becomes a rapidly oscillating function
in the limit of small piston velocities.} \label{fig:overlapplot}
\end{figure}

The prescription for evaluating the distribution is
\begin{equation}
P(W)=\frac{1}{\sqrt{2\pi}L}\int_0^L dx\int_{-\infty}^{\infty}dv
e^{-v^2/2}\delta \left( W-w_{\tau}(x,v) \right) \ .
\label{eq:distribution}
\end{equation}
The
calculations are presented in detail in appendix
\ref{sec:computations_for_P(W)}.  With the expression
(\ref{eq:expression_for_n}) for $n$, the number of bounces, one
can get rid of the summation over this number and the distribution
function simplifies to,
\begin{equation}
P(W) = \delta(W) P_{0} +
\frac{e^{-\frac{1}{2}\left(nv_p+\frac{W}{2nv_p}\right)^2}}{\sqrt{2\pi}nv_p}
f(W) \ . \label{dist}
\end{equation}
Here, $P_{0}$ is the probability to obtain vanishing work because
the molecule is unable to chase the piston or hit it even once,
\begin{equation} P_{0} = \frac{1}{\sqrt{2\pi}L}\int_0^Ldx\int_{-(L+v_p)}^{(L+v_p)}dv e^{-(v-x)^2/2} \ , \end{equation}
and the function $f(W)$, which we call the overlap factor, can be
formulated as follows (see also figure \ref{fig:overlapplot}):
\begin{equation} f(W) = \left\{ \begin{array}{lcr} -(n-1) \left(\frac{v_p}{2L}+1 \right) + \frac{W}{4nv_pL} & {\rm when} & (n-1)(v_p+2L) < \frac{W}{2nv_p} \leq (n-1)(v_p+2L)+2L \\
1 & {\rm when}  & (n-1)(v_p+2L)+2L < \frac{W}{2nv_p} \leq
(n-1)(v_p+2L)+2L+2v_p \\
(n+1)\left(\frac{v_p}{2L}+1 \right)-\frac{W}{4nv_pL}&{\rm when} &
(n-1)(v_p+2L)+2L+2v_p < \frac{W}{2nv_p} \leq (n+1)(v_p+2L)
\end{array} \right.\end{equation}
Here, the integer $n$ (which is the number of bounces by the
molecule against the piston) is obtained in appendix
\ref{sec:computations_for_P(W)} and is given by the formula
\begin{equation} n = \left[
\left.\left(1+\sqrt{1+\frac{2 W}{v_p(2L+v_p)}} \right) \right/ 2
\right] \ , \label{eq:expression_for_n}
\end{equation}
where $\left[ \ldots \right]$ means integer part of $\ldots$.  For
example, simple algebra indicates that as long as $W < 4 v_p(2 L +
v_p)$, we have just one collision, $n=1$.  For the values of work
$W$ in the next interval, $4 v_p(2 L +  v_p) < W < 12 v_p(2 L +
v_p) $, we have $n=2$, etc.

Thus, the probability distribution $P(W)$ consists of a
$\delta$-function peak at $W=0$ and a tail at positive $W$.

\subsection{Limit of large volume and fast moving piston}\label{sec:limit}

As we said in the beginning, the most interesting case is when the
piston moves fast, such that hardly any molecule can chase it and
produce non-zero work.  That means, the \J identity in this case
relies exclusively on the far tail of the Maxwell distribution. Let us
consider the probability distribution $P(W)$ in this limit, $v_p \gg
1$.

It is reasonable to assume simultaneously that the volume is large
enough, such that $L \gg v_p$.  Since in more traditional units
this condition reads $L \gg v_p \tau$, it means that the piston
moves fast, but for a very short time.

In this case, the distribution is dominated by the single bounce,
that is, $n=1$.  Assuming $n=1$ and $v_p \ll L$, we have
\begin{equation} f(W) \simeq \left\{ \begin{array}{lcr} \frac{W}{4v_pL} & {\rm when} & 0 < W < 4 v_p L \\
2 -\frac{W}{4v_pL}&{\rm when} & 4 v_p L < W < 8v_pL
\end{array} \right. \label{eq:f_approx}\end{equation}
which yields the probability distribution expression
\begin{equation}
P(W) \simeq \delta(W) P_{0} +
\frac{e^{-\frac{1}{2}\left(v_p+\frac{W}{2v_p}\right)^2}}{\sqrt{2\pi}v_p}
\frac{W}{4 v_pL} \ . \label{eq:dist1}
\end{equation}
valid at $L \gg v_p \gg 1$ and $W < 4 L v_p$.  For the larger $W$
tail of the distribution $P(W)$, we have to include the second
line of eq. (\ref{eq:f_approx}), and for even larger $W$ also
higher $n$ values.  Luckily, there is no need to do that, because the
simple approximation (\ref{eq:dist1}) is good enough to capture
the \J result.  Although the \J identity involves $\langle e^{W}
\rangle$, which includes integration over all values of $W$, the
integral converges rapidly enough to yield the correct answer within
the region of applicability of formula (\ref{eq:dist1}):
\begin{eqnarray}
 \langle e^W \rangle &=&\int_{-\infty}^{\infty} P(W)e^W dW= P_{0}+
\frac{1}{\sqrt{2\pi}v_p}\int_0^{\infty}e^{-\frac{1}{8v_p^2}(W-2v_p^2)^2}\times\frac{W}{4v_pL}dW\nonumber\\
&=&\int_0^{L}\frac{dx}{L}\frac{1}{\sqrt{2\pi}}\int_{-\infty}^{\infty}dv e^{-\frac{1}{2}(v-x)^2}+\nonumber\\
&&\frac{1}{\sqrt{2\pi}v_p}\int_{-\infty}^{\infty}e^{-\frac{1}{8v_p^2}(W-2v_p^2)^2}\times\frac{2v_p^2}{4v_pL}dW\nonumber\\
&=&1+\frac{v_p}{L} \ . \nonumber
\end{eqnarray}
Here, we made approximations in both the $P_0$ term, by extending the
integral limits to $(-\infty,\infty)$, and in the tail term, by
setting the integral lower limit to $-\infty$.  This way, we do
recover the \J formula (\ref{eq:Jarz1}).

Let us also calculate the probability of obtaining non-zero work
values (one or more collisions) as well as the average work done, i.e.
the $0^{th}$ (without the $P_0$ term) and $1^{st}$ moments of the distribution.
Using expression (\ref{eq:dist1}), we have
\begin{eqnarray}
P_{W>0} &=& \int_{0^+}^\infty P(W)dW = \frac{e^{-v_p^2/2}}{4\sqrt{2\pi}Lv_p^2}\int_0^\infty We^{-\frac{1}{2}W-\frac{1}{8v_p^2}W^2}dW \\
\left<W\right> &=& \int_{0}^\infty W P(W)dW = \frac{e^{-v_p^2/2}}{4\sqrt{2\pi}Lv_p^2}\int_0^\infty W^2e^{-\frac{1}{2}W-\frac{1}{8v_p^2}W^2}dW
\end{eqnarray}
For large $v_p$, we neglect the term $W^2/(8v_p^2)$ in the exponents upon integration, yielding
\begin{eqnarray}
P_{W>0} &\simeq& \frac{1}{\sqrt{2\pi}Lv_p^2}e^{-v_p^2 / 2} \label{problargevp}\\
\left<W\right> &\simeq& \frac{1}{\sqrt{2\pi}Lv_p^2}e^{-v_p^2 / 2}(4)=4P_{W>0} \label{averageworklargevp}
\end{eqnarray}

To the same approximation,
the probability to obtain zero work $W$ is equal to
\begin{equation} P_0 = 1-\frac{1}{\sqrt{2\pi}Lv_p^2}e^{-v_p^2 / 2} \
.
\end{equation}
As one could have expected, these probabilities are governed by
the tail of the Maxwell distribution.

\subsection{Comparison with simulations}

Measurements were made in computer simulations and compared
with the results obtained in the previous sections. The conditions
for the `pulling' experiments were as follows: the `pulling' time 
$\tau=1$, the temperature was set to $k_B T=1$ (which sets the width of
the gaussian distribution from which the initial velocities were selected),
and the number of trials or iterations used per measured average was $100,000$.
The parameters that we varied were the piston velocity $v_p$ and the initial
piston length (or `volume') $L$.

\begin{figure}
\centerline{\scalebox{1.0}{\includegraphics{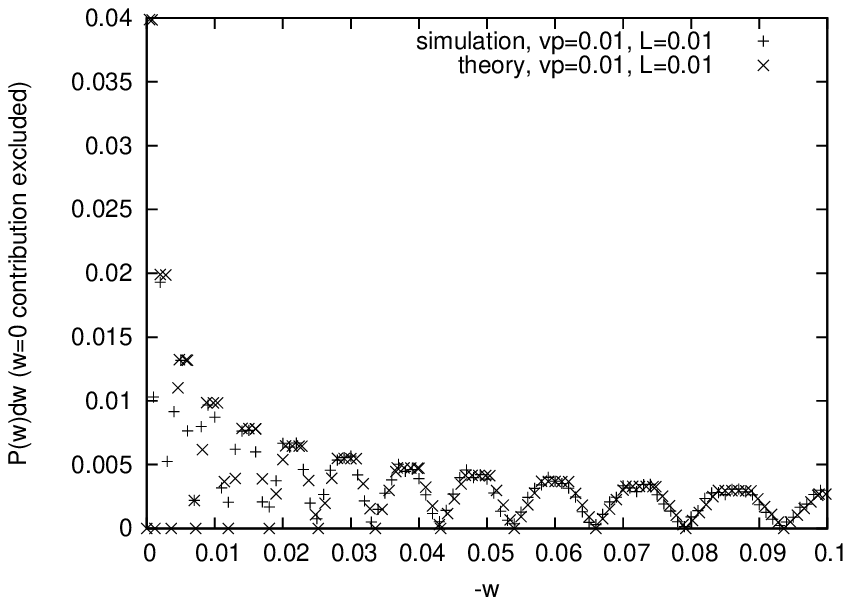}}}
\caption{Simulation results (pluses $+$) together with theoretical
calculation (expression (\ref{dist}), crosses $\times$) for the work distribution. For each
trial run, the cylinder volume was doubled ($v_p=0.01$, $\tau=1$,
$L=0.01$). The bin width used was $\Delta w=k_B T/1000=1/1000$.
The average number of collisions between molecule and
piston was about $20$.} \label{fig:pw01}
\end{figure}

\begin{figure}
\centerline{\scalebox{1.0}{\includegraphics{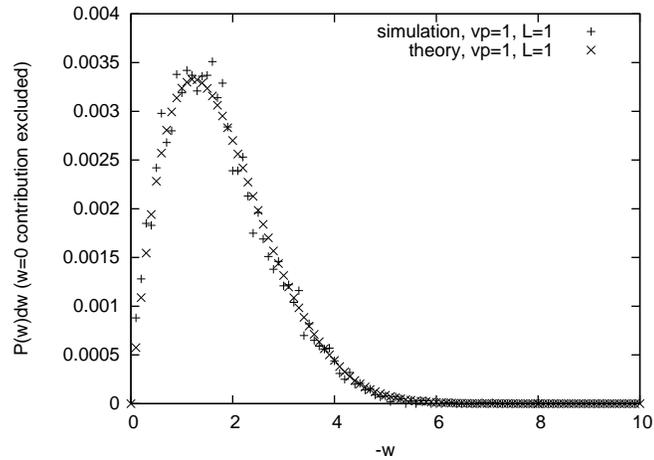}}}
\caption{Simulation results (pluses $+$) together with theoretical
calculation (crosses $\times$) for the work distribution. For each trial run, the cylinder
volume was doubled ($v_p=1$, $\tau=1$, $L=1$). 
The bin width used was $\Delta w=k_B T/10=1/10$.
The average number
of collisions between molecule and piston was low, about $0.08$.}
\label{fig:pw1}
\end{figure}

Figures \ref{fig:pw01} and \ref{fig:pw1} present the distribution of
probabilities
$P(W)dW$ for two different sets of piston velocities and piston lengths.
In the first case, the piston
velocity was set to $v_p=0.01$ and in the second case was set to $v_p=1$.
The effect of the overlap factor is evident in the former case of a 
slower moving piston.

\begin{figure}
\centerline{\scalebox{1.0}{\includegraphics{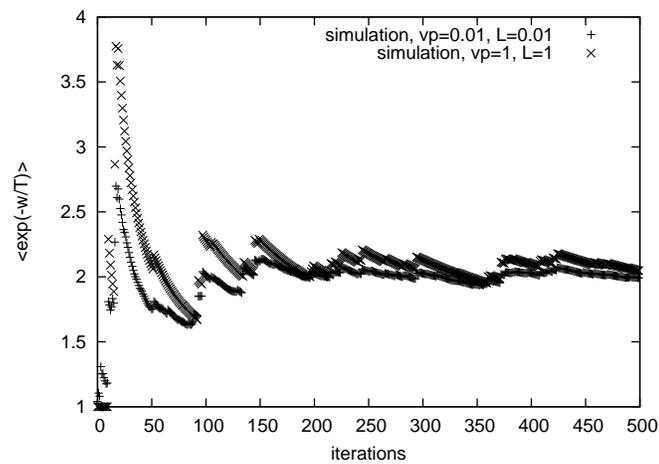}}}
\caption{Evolution of the average of $\exp(w/k_BT)$ with the number
of trials.} \label{fig:iters}
\end{figure}

Figure \ref{fig:iters} presents data for 
the average $\langle \exp(W/k_B T) \rangle$.
The \J identity predicts that this average should be
$(1+v_p/L)=2$.

\begin{figure}
\centerline{\scalebox{1.1}{\includegraphics{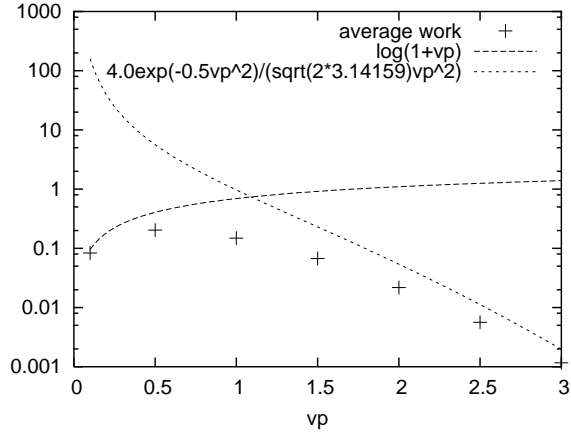}}}
\caption{Plot of measurements of the work done (pluses $+$)
together with the expression $\ln{(1+v_p\tau/L)}$ (i.e. `$-\Delta F$', with $\tau=1$ and $L=1$)
and the expression for the average work done for large $v_p$ (\ref{averageworklargevp}).} \label{fig:averagework}
\end{figure}

Figure \ref{fig:averagework} presents data for the average work done.
The expression corresponding to a free energy change at constant temperature
$T$, namely $\ln{(1+v_p/L)}=\ln\langle e^W \rangle=-\Delta F$, is plotted, as well as the expression
for the average work done (\ref{averageworklargevp}) for large $v_p$ ($v_p\gg 1$) .
As the velocity $v_p$ increases, the average work done is seen to shift from one regime (in which $\left<W\right>\simeq -\Delta F$) to another
(in which $\left<W\right> < -\Delta F$). If we take $-\left<W\right>-\Delta F=W_{\mbox{dis}}$ as some measure of `dissipation',
then it is also seen that this quantity increases as $v_p$ increases, although the difference is not
much more than $k_B T$.

\begin{figure}
\centerline{\scalebox{1.0}{\includegraphics{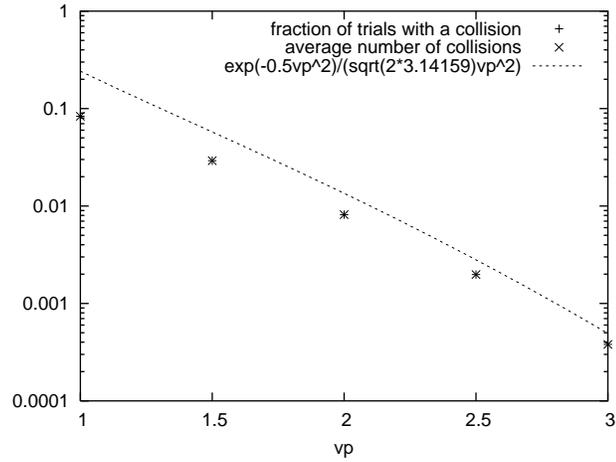}}}
\caption{Plot of the expression (\ref{problargevp})(line), the
fraction of trials with a collision (pluses $+$) and the average
number of collisions (crosses $\times$). The latter two
coincided.} \label{fig:prob}
\end{figure}

In figure \ref{fig:prob} the expression for the probability of obtaining non-zero work values in the high velocity limit 
is compared with the fraction of trials in which a collision occurred between molecule and piston.
Due to the rarity of collisions (and dominance of single collisions) in the range of velocities tested ($v_p=1,1.5,2,2.5,3$, $L=1$), the fraction of trials with collisions is identical to the average number of collisions.

\section{Discussion and conclusion}

Let us look closer at our main results obtained in section
\ref{sec:limit}.  One question to ask is this:  how many times
should one perform the experiment of moving the piston in order to
get a reasonable estimate of the average $\langle e^{W} \rangle$?
At the very least, in order to get the non-zero answer for the
free energy difference from \J formula (\ref{eq:Jarz}), one has to
get at least one case of non-zero work.  For this, one has to
perform about $1/P_{W>0}$ experiments, which is already a very
large number at $v_p \gg 1$.  In fact, as our calculations show,
in order to recover the \J identity, we have to continue
integration into the region where $W$ is as large as about $L
v_p$.  In practical terms, this means, we have to perform as many
runs on the system as to get at least a few realizations with the
work of this order.  According to the formula (\ref{eq:dist1}),
the corresponding probability is roughly proportional to $e^{-2
L^2}$.  In other words, this requires about $e^{+2 L^2}$ runs.
Restoring the more traditional notations with $k_B T$ and $\tau$,
we estimate the necessary number of runs (or trials) as $\exp
\left[ m L^2 / \tau^2 k_B T \right]$.  Clearly, this is a very
large number.

In practical terms, one may also want to know if the use of the \J identity is 
useful.  At first glance, it seems extremely useful: one apparently 
does not have to equilibrate the system and by doing purely 
non-equilibrium measurements, one nevertheless recovers the equilibrium free 
energy.  Our example suggests that the situation might be a little more 
tricky.  Indeed, to do equilibrium measurements, one has to proceed very 
slowly, to keep the system close to equilibrium all the time; for this, 
$\tau$ has to be larger than the system relaxation time, which grows 
with the system size $L$ 
(in our dimensionless variables, this corresponds to the limit $v_p
\ll L \ll 1$).  But, on the other hand, if one proceeds very 
rapidly, then one has to perform exponentially many experiments in order 
to catch the exponentially rare but decisively important fluctuations.  
This consideration suggests that there might be some optimal strategy.  
For the ideal gas model, such optimal strategy is most likely the 
(classical) slow `equilibrium' experiment, because the time for such 
an experiment grows only linearly with $L$, while the time for a `fast' 
experiment is exponential.  For other systems, the optimal strategy 
might be intermediate between one very slow experiment and very many  
rapid ones.  Unfortunately, it is clear that such optimal strategy is 
highly sensitive to the particularities of the system in question.  Our 
model, which is an ideal gas, is a system with a flat energy 
landscape.  For other energy landscapes one may wonder about the 
trajectories visiting various valleys.  Unfortunately, the knowledge of 
these valleys is exactly what one wants to learn from making measurements of 
equilibrium free energies.   In any case, the insight we can gain from our 
primitive model is that rapid non-equilibrium measurements are not 
automatically advantageous.

To conclude, we have presented a very naive simple model to look
at the \J identity.  We do recover the identity for our model, and
we are able to demonstrate that its validity relies on the far
tail of the Maxwell distribution, in the sense that the dominant
contribution is provided by the very rapidly moving molecules.  We
are also able to estimate how many independent experimental runs
are necessary to obtain the equilibrium free energy from the \J identity
with a reasonable accuracy, this necessary number of trials
appears exponential in the system size.

\section*{Acknowledgments}

Present work was initiated by the discussion during the Conference
on Statistical Physics of Macromolecules in Santa Fe in May of
2004, where one of the authors (AYG) formulated the ideal gas model studied here.
AYG acknowledges all participants of that discussion,
particularly D.Nelson, who pointed to the possible role of the tails of the Maxwell distribution in
the resolution of the paradox.  Computations for the present work were
performed using Minnesota Supercomputing Institute facilities.
This work was supported in part by the MRSEC Program of the
National Science Foundation under Award Number DMR-0212302.
                                                                                                                                                             
We are glad to present this paper to the journal honoring David
Chandler and his penetrating insight into the underlying
simplicity of complex physics.

\appendix

\section{Computing the numerator in Equation
\protect\ref{average}}\label{sec:calculating_I}

Using the expression for the work done in $n$ collisions
(\ref{eq:work_in_n}) and in view of the inequalities
(\ref{eq:ineq1}) and (\ref{eq:ineq2}) developed in section
\ref{sec:average}, we have
\begin{eqnarray}
I&=&\int_0^Ldx\int_{-\infty}^{\infty}dv e^{-\frac{1}{2}  v^2}e^{w_{\tau=1}(x,v)}\nonumber\\
&=&\int_0^Ldx\sum_{n=1}^{\infty}\int_{(2n-1)(L+v_p)-x}^{(2n+1)(L+v_p)-x}dv e^{-\frac{1}{2}  v^2}e^{- (-2vv_pn+2v_p^2n^2)}+\nonumber\\
&&\int_0^Ldx\sum_{n=1}^{\infty}\int_{(2n-1)(L+v_p)+x}^{(2n+1)(L+v_p)+x}dv e^{-\frac{1}{2}  v^2}e^{- (-2vv_pn+2v_p^2n^2)}+\nonumber\\
&&\int_0^Ldx\int_{-(L+v_p)-x}^{(L+v_p)-x}dv e^{-\frac{1}{2}  v^2}e^{0}\nonumber\\
&=&\int_0^Ldx\sum_{n=1}^{\infty}\left\{\int_{(2n-1)(L+v_p)-x}^{(2n+1)(L+v_p)-x}dv e^{-\frac{1}{2}  (v-2v_pn)^2}+\right.\nonumber\\
&&\left.\int_{(2n-1)(L+v_p)+x}^{(2n+1)(L+v_p)+x}dv e^{-\frac{1}{2}  (v-2v_pn)^2}\right\}+\nonumber\\
&&\int_0^Ldx\int_{-(L+v_p)-x}^{(L+v_p)-x}dv
e^{-\frac{1}{2}  v^2}\nonumber
\end{eqnarray}
Employing a change of variable, $v'=v-2nv_p$,
\begin{eqnarray}
I&=&\int_0^Ldx\sum_{n=1}^{\infty}\left\{\int_{(2n-1)L-x-v_p}^{(2n+1)L-x+v_p}dv' e^{-\frac{1}{2}  v'^2}+\right.\nonumber\\
&&\left.\int_{(2n-1)L+x-v_p}^{(2n+1)L+x+v_p}dv' e^{-\frac{1}{2}  v'^2}\right\}+\nonumber\\
&&\int_0^Ldx\int_{-(L+x)-v_p}^{L-x+v_p}dv e^{-\frac{1}{2}  v^2}\nonumber\\
\end{eqnarray}
If $v_p$ is zero or absent, the separate integrals could be
coalesced into a single integral of a gaussian from $-\infty$ to
$\infty$ and the result is trivial (identical to the denominator
in the average). Therefore, let us separate the `excess' from the
trivial result,
\begin{eqnarray}
I&=&\int_0^Ldx\int_{-\infty}^{\infty}dv e^{-\frac{1}{2}  v^2}+\nonumber\\
&&\sum_{n=1}^{\infty}\left\{\int_{(2n+1)L-x}^{(2n+1)L-x+v_p}dv+
 \int_{(2n-1)L-x-v_p}^{(2n-1)L-x}dv+\right.\nonumber\\
&&\left.\int_{(2n+1)L+x}^{(2n+1)L+x+v_p}dv+
 \int_{(2n-1)L+x-v_p}^{(2n-1)L+x}dv\right\}e^{-\frac{1}{2}  v^2}+\nonumber\\
&&\left\{\int_{L-x}^{L-x+v_p}dv+
 \int_{L+x}^{L+x+v_p}dv\right\}e^{-\frac{1}{2}  v^2}\nonumber
\end{eqnarray}
The first, third, fifth and sixth integrals after the summation
symbol (those with upper limits ``$...+v_p$'') can be combined
after making the change of variables, $x'=(2n+1)L-x$,
$x'=(2n+1)L+x$, $x'=L-x$, $x'=L+x$, yielding
\begin{equation}
I_1=\int_0^\infty dx'\int_{x'}^{x'+v_p}dv
e^{-\frac{1}{2}  v^2}
\end{equation}
Combining the second and fourth integrals after the summation
symbol (those with lower limits ``$...-v_p$'') in a similar away,
after the change of variables $x'=(2n-1)L-x$, $x'=(2n-1)L+x$, the
result is
\begin{equation}
I_2=\int_0^\infty dx'\int_{x'-v_p}^{x'}dv
e^{-\frac{1}{2}  v^2}
\end{equation}
Performing yet another change of variables, $v'=v-x'$,
\begin{equation}
I_1=\int_0^\infty dx'\int_{0}^{v_p}dv' e^{-\frac{1}{2}
(v'+x')^2}
\end{equation}
\begin{equation}
I_2=\int_0^\infty dx'\int_{-v_p}^{0}dv' e^{-\frac{1}{2}
(v'+x')^2}
\end{equation}
(Notice that the series of variable substitutions effectively
exchanged the infinite limits associated with $v$ with the finite
limits associated with $x$.) After performing the change of
variable for $I_2$, $v'=-v''$, and combining the two integrals,
\begin{eqnarray}
I_1+I_2&=&\int_0^\infty dx'\int_{0}^{v_p}dv' e^{-\frac{1}{2}  (x'+v')^2}+\nonumber\\
&&\int_0^\infty dx'\int_{0}^{v_p}dv'' e^{-\frac{1}{2}  (x'-v'')^2}\nonumber\\
&=&\int_0^{v_p} dv'\int_{-\infty}^{\infty}dx' e^{-\frac{1}{2}
(x'-v')^2}\nonumber
\end{eqnarray}
Exchanging the ``roles'' of $x'$ and $v'$, i.e. letting
$v=x'$ and $x=v'$,
\begin{equation}
I_1+I_2=\int_0^{v_p}dx\int_{-\infty}^{\infty}dv
e^{-\frac{1}{2}  v^2}
\end{equation}

Using this simplified result, the numerator in the average becomes
\begin{eqnarray}
I&=&\int_0^{L}dx\int_{-\infty}^{\infty}dv e^{-\frac{1}{2} v^2}
+\int_0^{v_p}dx\int_{-\infty}^{\infty}dv e^{-\frac{1}{2}  v^2}\nonumber\\
&=&\int_0^{L+v_p}dx\int_{-\infty}^{\infty}dv e^{-\frac{1}{2}
v^2} \ \label{eq:numerator}
\end{eqnarray}

\section{Computing the probability distribution $P(W)$, Eq. (\protect\ref{eq:distribution})}\label{sec:computations_for_P(W)}

The inequalities (\ref{eq:ineq1}) and (\ref{eq:ineq2}) developed
in section \ref{sec:average} lead to the following partition of
the integral
\begin{eqnarray}
P(W)&=& \frac{1}{\sqrt{2\pi}L} \int_0^Ldx\sum_{n=1}^{\infty} \int_{(2n-1)(L+v_p)-x}^{(2n+1)(L+v_p)-x}dv e^{-v^2/2} \times \delta \left( W- (2vv_pn-2v_p^2n^2) \right) + \nonumber\\
&&\frac{1}{\sqrt{2\pi}L}\int_0^Ldx\sum_{n=1}^{\infty} \int_{(2n-1)(L+v_p)+x}^{(2n+1)(L+v_p)+x}dv e^{-v^2/2} \times \delta \left( W-(2vv_pn-2v_p^2n^2) \right) + \nonumber\\
&&\frac{1}{\sqrt{2\pi}L}\int_0^Ldx\int_{-(L+v_p)-x}^{(L+v_p)-x}dv
e^{-v^2/2}\delta(W-0)\nonumber
\end{eqnarray}

Call the first term $I_1$ and the second term $I_2$ (the third
term is `trivial'). Performing a change of variable to remove $x$
from the limits, integrating over $x$ and taking advantage of the
delta function results in
\begin{eqnarray}
I_1&=&\sum_{n=1}^{\infty}\frac{e^{-\frac{1}{2}\left(nv_p+\frac{W}{2nv_p}\right)^2}}{\sqrt{2\pi}nv_p}\times\frac{1}{2L}\left\{\mbox{overlap between}\right.\nonumber\\
&&\left.\left[nv_p+\frac{W}{2nv_p},nv_p+\frac{W}{2nv_p}+L\right]\right. \nonumber\\
&&\left.\mbox{and}\left[(2n-1)(L+v_p),(2n+1)(L+v_p)\right]
\right\}\nonumber
\end{eqnarray}
and similarly for $I_2$,
\begin{eqnarray}
I_2&=&\sum_{n=1}^{\infty}\frac{e^{-\frac{1}{2}\left(nv_p+\frac{W}{2nv_p}\right)^2}}{\sqrt{2\pi}nv_p}\times\frac{1}{2L}\left\{\mbox{overlap between}\right.\nonumber\\
&&\left.\left[nv_p+\frac{W}{2nv_p}-L,nv_p+\frac{W}{2nv_p}\right]\right. \nonumber\\
&&\left.\mbox{and}\left[(2n-1)(L+v_p),(2n+1)(L+v_p)\right]
\right\}\nonumber
\end{eqnarray}
$I_1$ and $I_2$ can be combined as follows,
\begin{eqnarray}
I_1+I_2&=&\sum_{n=1}^{\infty}\frac{e^{-\frac{1}{2}\left(nv_p+\frac{W}{2nv_p}\right)^2}}{\sqrt{2\pi}nv_p}\times\frac{1}{2L}\left\{\mbox{overlap between}\right.\nonumber\\
&&\left.\left[nv_p+\frac{W}{2nv_p}-L,nv_p+\frac{W}{2nv_p}+L\right]\right. \nonumber\\
&&\left.\mbox{and}\left[(2n-1)(L+v_p),(2n+1)(L+v_p)\right] \right\}\nonumber\\
&=&\sum_{n=1}^{\infty}\frac{e^{-\frac{1}{2}\left(nv_p+\frac{W}{2nv_p}\right)^2}}{\sqrt{2\pi}nv_p}\times
f(n,W) \nonumber
\end{eqnarray}
where the overlap factor $f$ satisfies $0 \leq f \leq 1$, since
the range of the smaller interval is at most $2L$. $f$ is also
zero for negative $W$, or positive work values $-W$ done by the piston.

The conditions that must be satisfied by $W$ in order for the
overlap associated with integer $n$ to occur are
\begin{eqnarray}
2nv_p\left(2(n-1)L+(n-1)v_p\right)<W<2nv_p\left(2(n+1)L+(n+1)v_p\right)
\label{wint}
\end{eqnarray}
Notice that the left boundary of the interval is a function of
$n(n-1)$, while the right interval is a function of $n(n+1)$.
Therefore the right boundary can be transformed into the
left-boundary by making the replacement $n\rightarrow n-1$. This
implies that the intervals (\ref{wint}) are contiguous and
nonoverlapping and that at most one term in the summation in
$P(W)$ survives. One can solve for the integer $n$ by taking the
integer part (or floor function) of a solution to a quadratic
equation,
\begin{equation}
W > 2nv_p\left(2(n+1)L+(n+1)v_p \right)
\end{equation}
which results in formula (\ref{eq:expression_for_n}) in the main
text.

\end{document}